\documentclass[%
reprint,
superscriptaddress,
 amsmath,amssymb,
 aps,
]{revtex4-2}

\usepackage[utf8]{inputenc}
\usepackage{graphicx}
\usepackage{dcolumn}
\usepackage{multirow}
\usepackage{bm}
\usepackage{mathrsfs}
\usepackage{hyperref}
\hypersetup{
  colorlinks = true,
  urlcolor = blue,
  linkcolor = blue,
  citecolor = green,
  filecolor = magenta,
}

\newcommand{\ud}{\mathrm{d}}
\newcommand{\pd}{\partial}

\newcommand{\mm}{\mathfrak m}

\begin{document}


\title{Dark photon bursts from compact binary systems and constraints}

\author{Shaoqi Hou}
\email{hou.shaoqi@whu.edu.cn}
\affiliation{School of Physics and Technology, Wuhan University, Wuhan, Hubei 430072, China}
\author{Shuxun Tian}
\affiliation{Department of Astronomy, Beijing Normal University, Beijing 100875,  China}
\author{Shuo Cao}
\affiliation{Department of Astronomy, Beijing Normal University, Beijing 100875,  China}
\author{Zong-Hong Zhu}
\email{zhuzh@whu.edu.cn}
\affiliation{School of Physics and Technology, Wuhan University, Wuhan, Hubei 430072, China}
\affiliation{Department of Astronomy, Beijing Normal University, Beijing 100875,  China}

\date{\today}

\begin{abstract}
   In this work, we consider the burst signal of the dark photon, the hypothetical vector boson of the $U(1)_B$ or $U(1)_{B-L}$ gauge group, generated by a compact binary star system. 
   The absence of the signal in the laser interferometer puts bounds on the coupling constant $\epsilon$ to the ordinary matter.
   It turns out that if the dark photon is massless, $\epsilon^2$ is on the order of $10^{-37}-10^{-33}$ at most; in the massive case, the upper bound of $\epsilon^2$ is about $10^{-38}-10^{-31}$ in the mass range from $10^{-19}$ eV to $10^{-11}$ eV. 
   These are the first bounds derived from the interferometer observations independent of the assumption of dark photons being dark matter.
\end{abstract}

\maketitle


\section{Introduction}

The excellent precision of the laser interferometer not only made it possible to detect gravitational waves (GWs) \cite{Abbott2016.PRL.116.061102,Abbott2017.PRL.119.161101,Abbott2021.ApJL.915.L5}, but also enables the observation of new physics, such as the quantum mechanics of macroscopic objects \cite{LIGOScientific:2020luc,Whittle:2021mtt}.
Other new physics includes new elementary particles as candidates for dark matter (DM). 
One of the possibilities is the dark photon (DP), which is the gauge boson associated with $U(1)_B$ or $U(1)_{B-L}$, and whose mass $m_{\gamma'}$ can be generated via the St\"ueckelberg mechanism \cite{Ruegg:2003ps}.
Since ordinary matter, such as the mirrors in interferometers, usually are charged under these groups, they can be accelerated by DPs.
Thus, interferometers are capable of detecting DPs.
In Refs.~\cite{Pierce2018.PRL.121.061102,Guo:2019ker,Morisaki2021.PRD.103.L051702,LIGOScientific:2021odm}, the ultralight DP was considered, and the stochastic background of DPs is thus coherently oscillating.
Using the technique of detecting the stochastic GW background by interferometers \cite{Callister:2017ocg,Abbott:2018utx}, DP model was highly constrained.

Here, we continue to test the DP model with GW interferometers.
Instead of assuming that DPs constitute DM and using the observed DM density, we will study the emission of DPs by orbiting compact stars, just like the emission of ordinary photons by electrically charged particles that are accelerated.
The generated DPs reach the interferometer, and similarly to GWs, cause strain which can be measured.
Since the observed GW waveforms agree with predictions of general relativity (GR)  very well, the DP model can thus be bounded.

Very interestingly, the induced strain by DPs is an explicit function of the source redshift in the frequency domain, if DP is massive.
This enables the measurement of the redshift directly with DP radiation.
Together with the luminosity distance determined with GWs (the idea of standard sirens \cite{Schutz:1986gp}), the measurement of DP radiation with the laser interferometer may play important roles in cosmology.
Of course, the DP has not been detected. 
But our result may stimulate the search for suitable matter field radiation like DPs to determine the redshift.

Besides the methods presented  in Refs.~\cite{Pierce2018.PRL.121.061102,Guo:2019ker,Morisaki2021.PRD.103.L051702,LIGOScientific:2021odm} and in the current work, one can also constrain DP parameters via the measurement of the violation of the equivalence principle \cite{Berge:2017ovy,Schlamminger:2007ht}.
Indeed, as long as objects carry different ratios of ``dark charge'' to mass, they accelerate differently in the same uniform ``electric'' field of the DP, so the universal free-fall is violated \cite{Will:1993ns}.
The study of black hole superradiance might also provide strong constraints \cite{Arvanitaki:2014wva,Baryakhtar:2017ngi,East:2017ovw,East:2017mrj,Cardoso:2018tly,Caputo:2021efm}, just like the superradiance of scalar particles \cite{Brito:2015oca}.
In the end, the charged stars under $U(1)_B$ or $U(1)_{B-L}$ may change their positions due to the DP radiation, so Gaia mission and the like are capable of detecting DPs \cite{Guo:2019qgs}.

In fact, the idea of extra $U(1)$ groups has a long history, e.g., \cite{Fayet:1980ad,*Fayet:1980rr,*Fayet:1989mq,*Fayet:1990wx}, where the new vector boson is called U-boson.
They might originate from Grand Unification, (super)string theory, higher dimensional theories, etc..
U-bosons can be made massive via the Higgs mechanism, assuming there exists one extra Higgs boson which is singlet under the standard model gauge transformation but charged under the new $U(1)$.
This new $U(1)$ charge is generally proportional to a linear combination of the baryon and lepton numbers for the electrically charged neutral object, and within Grand Unification, the charge is simply proportional to $B-L$.
The presence of the massive U-bosons modifies the central force between binary stars by the Yukawa force, the fifth force, which can be tested by MICROSCOPE Mission \cite{Berge:2017ovy,Fayet:2017pdp,*Fayet:2018cjy}.
Cardoso \textit{et.~al.} also considered the emission of the vector boson of a hidden $U(1)$ symmetry by binary systems \cite{Cardoso:2016olt}. 
The emission carries more energy of the binary system away, and modifies the phase evolution of the GW. 
This can be used to test the vector boson model.
In the current work, the presence of DPs is inferred directly from the strain induced by them, instead of from the GW strain.
The motion of the binary stars with the electric and magnetic charges is also investigated in Refs.~\cite{Liu:2020vsy,*Liu:2020bag}, where more complicated orbits exist and the attention was still on the GW waveform.
One can also consider the $U(1)$ group of the lepton number differences, and Refs.~\cite{KumarPoddar:2019ceq,*KumarPoddar:2020kdz,Dror:2019uea} constrained such models using the observations on the orbital decay of binaries and perihelion precession.

This work is organized as follows. 
Section~\ref{sec-bdp} reviews the basics of DPs, especially the equations of motion and the energy-momentum tensor that are useful for computing DP radiation and the power carried away.
Based on this, the DP radiation is computed in Sec.~\ref{sec-dp-rad}.
There, one first notices the changes in the central force due to new $U(1)_B$ or $U(1)_{B-L}$ interaction in Sec.~\ref{sec-mcf}. 
Then, one can calculate the ``electric'' dipole radiation following the technique learned in Ref.~\cite{Jackson:1998nia} in Sec.~\ref{sec-edp}, followed by the discussion on ``magnetic'' dipole and ``electric'' quadrupole radiation in Sec.~\ref{sec-mdp-eqp}.
Section~\ref{sec-res} focuses on how the DP radiation induces strain in the laser interferometer, and by requiring the signal-to-noise ratio of this strain be small enough, the constraints can be obtained, as presented in Sec.~\ref{sec-cts}.
Section~\ref{sec-con} concludes this work.

\section{The basics of dark photons}
\label{sec-bdp}

The action of the DP $A^\mu=(V/c,\vec A)$ takes that of the Proca field \cite{Jackson:1998nia},
\begin{equation}
    \label{eq-proca}
    \mathscr L=-\frac{1}{4\mu_0}F_{\mu\nu}F^{\mu\nu}+\frac{\mathfrak m^2}{2\mu_0}A_\mu A^\mu-\epsilon eJ_\mu A^\mu,
\end{equation}
where $F_{\mu\nu}=\pd_\mu A_\nu-\pd_\nu A_\mu$ is the field strength, $\epsilon$ the DP coupling constant with the matter, $e$ the electric charge, $\mu_0$ the permeability for vacuum, $\mm=m_{\gamma'}c/\hbar$ the inverse Compton wavelength, and $J^\mu=(c\rho,\vec j)$ the $B$ or $B-L$ number flux.
The mostly minus sign convention for the metric is used.
The equations of motion are given by
\begin{gather}
    \pd^\nu\pd\nu A_{\mu}+\mathfrak m^2 A_\mu=\mu_0\epsilon eJ_\mu,\label{eq-eoms}\\
    \pd_\mu A^\mu=0,\label{eq-cs}
\end{gather}
where the second equation becomes a gauge condition in the massless case.
This set of equations can also be written in the following equivalent form,
\begin{gather}
\nabla\cdot\vec E+\mathfrak m^2V=\frac{\epsilon e\rho}{\epsilon_0},\\
\nabla\times\vec H-\epsilon_0\frac{\partial}{\partial t}\vec E+\frac{\mathfrak m^2}{\mu_0}\vec A=\epsilon e\vec j,\label{eq-4efh}\\
\nabla\cdot\vec H=0,\quad\nabla\times\vec E+\mu_0\frac{\partial}{\partial t}\vec H=0,\\
\nabla\cdot\vec A+\frac{1}{c^2}\frac{\partial}{\partial t}V=0,\label{eq-cs-gf}
\end{gather}
where $\epsilon_0$ is the dielectric constant of vacuum, and $\vec E=-\nabla V-\pd\vec A/\pd t$ and $\vec H=\nabla\times\vec A/\mu_0$ are ``electric'' and ``magnetic'' fields for the DP.

The test particles, such as the mirrors in the interferometer, would be accelerated if there exists nontrivial DP field with the following 3-acceleration, 
\begin{equation}
    \label{eq-lf}
    \vec a=\frac{\epsilon eN}{m}(\vec E+\vec v\times\vec B),
\end{equation}
with $\vec B=\mu_0\vec H$.
Also,  $N=\sigma m\chi/m_p+m\chi'/m_n$ is $B$ (if $\sigma=1$) or $B-L$ (if $\sigma=0$) number, where $m_p$ and $m_n$ are proton and neutron masses, $\chi$ and $\chi'$ are the mass fractions for protons and neutrons, respectively.
This enables the use of the GW interferometers to detect the DP produced by the binary system.

Finally, the stress-energy tensor can be easily obtained by the variation of the action with respect to $g^{\mu\nu}$, given by \cite{Jackson:1998nia}
\begin{equation*}
    \label{eq-set}
    \begin{split}
    T_{\mu\nu}=\frac{1}{\mu_0}&\left[ F_{\mu\rho} F^{\rho}{}_\nu+\frac{1}{4}g_{\mu\nu}F_{\rho\sigma}F^{\rho\sigma}\right.\\
    &\left.+\mm^2 \left( A_\mu A_\nu -\frac{1}{2}g_{\mu\nu}A_\rho A^\rho\right) \right].
    \end{split}
\end{equation*}
The temporal-spatial components are useful for computing the energy flux density, which is
\begin{equation*}
    \label{eq-set-0j}
    T^{0j}=\frac{1}{c}\left( \vec E\times\vec H+\frac{\mm^2}{\mu_0}\Phi\vec A \right)^j,
\end{equation*}
so one can recognize the Poynting vector 
\begin{equation*}
    \label{eq-def-poy}
    \vec S_m=\vec E\times\vec H+\frac{\mm^2}{\mu_0}\Phi\vec A,
\end{equation*}
where the first term is the Poynting vector in the massless case \cite{Jackson:1998nia}.
Since in our calculation, we treat all fields complex functions, the Poynting vector is 
\begin{equation}
    \label{eq-poy-c}
\vec S_m=\frac{1}{2}\Re\left(\vec E\times\vec H^*+\frac{\mm^2}{\mu_0}V\vec A^*\right),
\end{equation}
where $\Re$ means to take the real part.
This expression is used to calculate the radiated energy by the binary system.

Note that one assumes the flat spacetime background in writing down the above expressions.
Although in the vicinity of the binary system, the spacetime is curved, the curvature is small.
The perturbations to the dark photon radiation due to the curvature will be of the higher orders, so we will ignore them.
With these equations, one can compute the DP radiation produced by a binary system.

\section{Dark photon radiation from the binary system}
\label{sec-dp-rad}

In the classical electrodynamics, two opposite charges orbiting around each other radiate light \cite{Jackson:1998nia}.
Similarly, two orbiting stars may also emit DPs. 
In this section, we will consider DP radiation up to the ``electric'' quadrupole order, just like the ordinary photon radiation.

\subsection{Modified central force}
\label{sec-mcf}

Since one considers a new $U(1)$ symmetry \footnote{Either $U(1)_B$ or $U(1)_{B-L}$}, the orbital motion of binary stars would be modified becomes of the new interaction induced by the DP field strength $(\vec E,\vec B)$.
By the temporal component of Eq.~\eqref{eq-eoms}  in the static case, the scalar potential is  the Yukawa potential,
\begin{equation}
   \label{eq-vp-st} 
V=\frac{\epsilon eN}{4\pi\epsilon_0r}e^{-\mathfrak mr},
\end{equation}
due to a star with the new $U(1)$-charge $eN$.
We will consider the use of interferometers to detect DPs, so $10^{-19}\text{ eV}<m_{\gamma'}c^2<10^{-11}\text{ eV}$, corresponding to the Compton wavelength in the range of ($10^4\text{ m},\,10^{12}\text{ m}$), which is much smaller than the distances between stars in a binary system at least in the inspiral stage. 
Therefore, one can ignore the exponential factor in describing the motion of the stars with masses $m_1$ and $m_2$, that is, the acceleration of the effective one body problem is \cite{Cardoso:2016olt}
\begin{equation}
    \label{eq-eob-a}
    r\omega^2=G\frac{M}{r^2}-\frac{\epsilon^2e^2}{4\pi\epsilon_0}\frac{N_1N_2}{r^2}\frac{M}{m_1m_2}=G'\frac{M}{r^2},
\end{equation}
with $M=m_1+m_2$.
One can thus define an effective gravitational constant $G'=G-\frac{\epsilon^2e^2}{4\pi\epsilon_0}\zeta_1\zeta_2$ with $\zeta=N/m=\sigma\chi/m_p+\chi'/m_n$.

Of course, $G'$ should be very close to $G$, otherwise the orbital motion of binary stars would be modified so much that one can detect the difference easily, which has not happened yet \cite{Hulse:1974eb,Abbott:2018lct,LIGOScientific:2019fpa}.
So one can require that $G'\approx G$ to bound $\epsilon^2$.
In this work, we choose the binary systems in Table~\ref{tab-sp} to constrain $\epsilon^2$.
The first three sources \cite{LIGOScientific:2018mvr,LIGOScientific:2021qlt} can emit DPs observable by the ground-based interferometers  like aLIGO, Einstein Telescope (ET) \cite{Punturo:2010zza} and Cosmic Explorer (CE) \cite{Evans:2016mbw}.
They can also be observed by the DECihertz laser Interferometer Gravitational wave Observatory (DECIGO) and its downscale version, B-DECIGO \cite{Seto:2001qf,Isoyama:2018rjb}. 
The remaining four sources are for LISA \cite{Seoane:2013qna}, where EMRI stands for extreme mass-ratio inspiral, IMRI intermediate mass-ratio inspiral, IMBH intermediate mass black hole binary, and SMBH supermassive black hole binary \cite{Chamberlain:2017fjl}. 
One also assumes that the mass fraction $\chi$ of protons in a black hole is approximately 0.5, and $\chi=0$ for a neutron star.
\begin{table}
    \centering
    \begin{tabular}{c|ccc}
        \hline\hline
        Binary & $m_1(M_\odot)$ & $m_2(M_\odot)$ & $z$\\
        \hline
        GW150914 & 35.6 & 30.6 & 0.09\\
        GW200105 & 8.9 & 1.9 & 0.06 \\
        GW170817 & 1.46 & 1.27 & 0.01 \\
        \hline
        EMRI & $10^5$ & 10 & 0.2\\
        IMRI & $10^5$ & $10^3$ & 0.8\\
        IMBH & $5\times10^3$ & $4\times10^3$ & 2.0\\
        SMBH & $5\times10^6$ & $4\times10^6$ & 5.0\\
        \hline\hline
    \end{tabular}
    \caption{Source properties.
    The first 3 rows are taken from Refs.~\cite{LIGOScientific:2018mvr,LIGOScientific:2021qlt}.
    The remaining 4 rows are suggested by Ref.~\cite{Chamberlain:2017fjl}, where EMRI stands for extreme mass-ratio inspiral, IMRI stands for intermediate mass-ratio inspiral, IMBH intermediate mass black hole binary, and SMBH supermassive black hole binary.}
    \label{tab-sp}
\end{table}
Then, one finds out that 
\begin{equation}
    \label{eq-cs-or}
\epsilon^2\lesssim(8.1-32.5)\times 10^{-37}
\end{equation}
for both gauge groups, independent of the mass $m_{\gamma'}$.

Here, as listed in Table~\ref{tab-sp}, there are several binary black hole systems.
Whether black holes still carry $U(1)_B$ or $U(1)_{B-L}$ charges long after their formation is an interesting question. 
It is known that black holes with the standard model electric charge can discharge due to vacuum polarization, but this process is very slow \cite{Gibbons:1975kk}.
For $U(1)_B$ or $U(1)_{B-L}$ charges, the time scale of discharge is on the order of $\hbar/m_{\gamma'}c^2$ \cite{Coleman:1991ku}.
Since ultralight DPs are considered in the present work, black holes may still carry some amount of such dark charges after the formation.
There is another point worth to be mentioned. 
Dark matter may also accumulate around supermassive black holes \cite{Quinlan:1994ed,Bertone:2005hw,Zhao:2005zr}, and they can be charged under $U(1)_B$ or $U(1)_{B-L}$. 
As black holes are circling around in their orbits, dark matter would also be dragged by the gravitational pull of black holes, and effectively, black holes are charged.
This is also an interesting topic to consider, but beyond the scope of our current work.
So here, we will assume the black hole carries $U(1)_B$ or $U(1)_{B-L}$ charge, following Ref.~\cite{Dror:2021wrl}.

\subsection{``Electric'' dipole radiation}
\label{sec-edp}

Now, consider the DP radiation.
The solution to the spatial components of Eq.~\eqref{eq-eoms} is generally given by
\begin{equation}
    \label{eq-vp-1}
    \begin{split}
    \vec A(t,\vec x)=&\frac{\mu_0\epsilon e}{4\pi}\Bigg[\int\ud^3x'\frac{\vec j(t-|\vec x-\vec x'|/c,\vec x')}{|\vec x-\vec x'|}\\
    &-\int c\ud t'\ud^3x'\Theta(\Delta|x|)
        \frac{\mathfrak m^2J_1(\zeta)}{\zeta}\vec j(t',\vec x')\Bigg],
    \end{split}
\end{equation}
where $\zeta=\mathfrak m\sqrt{c^2(t-t')^2-|\vec x-\vec x'|^2}$ and $\Delta|x|=c(t-t')-|\vec x-\vec x'|$.
Note that in the massive case, the Green function is \cite{Poisson:2011nh}
\begin{equation*}
    \label{eq-def-gf-ms}
    G(ct,\vec x)=\frac{\delta(ct-|\vec x|)}{|\vec x|}-\Theta(ct-|\vec x|)\frac{\mathfrak mJ_1(\mathfrak m\sqrt{c^2t^2-\vec x^2})}{\sqrt{c^2t^2-\vec x^2}},
\end{equation*}
where  $J_1(z)$ is the Bessel function of the first kind.
Since $J_1(0)=0$, this Green function smoothly reduces to the one for the massless DP.
So in the massless case, $\vec A(t,\vec x)$ is given by the first integral in Eq.~\eqref{eq-vp-1}.

In the radiation zone, one can approximate $|\vec x-\vec x'|\approx R-\hat n\cdot\vec x'$ with $R=|\vec x|$ and $\hat n=\vec x/R$.
If one assumes the density $\rho(t,\vec x)=\rho(\vec x)e^{-i\omega t}$ and the 3-current $\vec j(t,\vec x)=\vec j(\vec x)e^{-i\omega t}$, one obtains the following approximation \cite{Jackson:1998nia,Alsing:2011er}
\begin{equation*}
    \label{eq-4pot}
    \begin{split}
    \vec A(\vec x)\approx &\frac{\mu_0\epsilon e}{4\pi}\frac{e^{i\omega R/c}}{R}\int\ud^3x'\vec j(\vec x')(1-ik\hat n\cdot\vec x')\\
    &-\frac{\mu_0\epsilon e}{4\pi R}\int\ud^3x'\int_0^\infty\ud\zeta J_1(\zeta)\vec j(\vec x')\times\\
    &e^{i\frac{\omega}{c}R\sqrt{1+(\zeta/\mathfrak mR)^2}} \left[ 1-i\frac{\omega}{c}\frac{\hat n\cdot\vec x'}{\sqrt{1+(\zeta/\mathfrak mR)^2}} \right].
    \end{split}
\end{equation*}
Here, we have omitted a factor of $e^{-i\omega t}$.
Once $\vec A$ is determined, one knows that  
\begin{gather}
    V(\vec x)=-i\frac{c^2}{\omega}\nabla\cdot\vec A(\vec x),\\
    \vec H=\frac{1}{\mu_0}\nabla\times\vec A,\\
    \vec E=\frac{i}{\omega\epsilon_0}\nabla\times\vec H+\frac{i}{\omega}\mathfrak m^2c^2\vec A.
\end{gather}
The first equation is due to Eq.~\eqref{eq-cs-gf}, and the last one due to Eq.~\eqref{eq-4efh}.
Following Ref.~\cite{Jackson:1998nia}, one introduces ``electric'' dipole, ``magnetic'' dipole, and ``electric'' quadrupole moments,
\begin{gather}
    \vec D_e=\int\vec x\rho(\vec x)\ud^3x,\quad \vec D_m=\frac{1}{2}\int\vec x\times\vec j(\vec x)\ud^3x,\\
    Q_{jk}=\int3x_jx_k\rho(\vec x)\ud^3x,
\end{gather}
respectively.
Their variations lead to the emission of DPs.

Let us consider first the field strength related to the ``electric'' dipole moment.
Assuming the binary stars  move around each other in a quasi-circular orbit in the $xOy$ plane, one can check that 
\begin{gather}
    \vec D_e=d_e\hat{\mathscr E} ,\; d_e=\eta^{8/15}\mathcal M\left( \frac{G'\mathcal M}{\omega^2} \right)^{1/3}\xi_e,\nonumber\\
    \hat{\mathscr E}=(1,i,0),\quad\xi_e=\sigma\frac{\Delta \chi}{ m_p}+\frac{\Delta \chi'}{m_n},\label{eq-eds}
\end{gather}
where $\eta=m_1m_2/M^2$ is the symmetric mass ratio, and $\mathcal M=\eta^{3/5}M$ is the chirp mass. 
In addition, $\Delta \chi=\chi_1-\chi_2$ with $\chi_1$ and $\chi_2$ the proton mass fractions of the two stars, and $\Delta \chi'=\chi'_1-\chi'_2$ with $\chi'_1$ and $\chi'_2$ the neutron mass fractions.

The variation of $\vec D_e$  sources 
\begin{equation}
    \label{eq-ed-rad}
    \vec A_\text{ed}=-\frac{i\mu_0\epsilon e\omega}{4\pi R}\left[e^{i\omega R/c}-I_0^{(\omega)}(R)\right]\vec D_e,
\end{equation}
where the subscript means ``electric dipole'', and we have defined the symbol,
\begin{equation}
    \label{eq-def-in}
    I_n^{(\omega)}(R)=\int_0^\infty\ud\zeta J_1(\zeta)\frac{e^{ i\frac{\omega}{c}R\sqrt{1+( \zeta/\mathfrak mR )^2} }}{\left[ 1+(\zeta/\mathfrak mR)^2 \right]^{n/2}}.
\end{equation}
In the radiation zone, it is known that \cite{Alsing:2011er}
\begin{equation}
    \label{eq-inapp}
    I_n^{(\omega)}\approx e^{i\omega R/c}-\left[ 1-\left( \frac{\mathfrak mc}{\omega} \right)^2 \right]^{\frac{n-1}{2}}e^{ i\frac{\omega R}{c} \sqrt{1-\left( \frac{\mathfrak mc}{\omega} \right)^2}},
\end{equation}
when $\omega>\mathfrak mc$; otherwise, $I_n^{(\omega)}\approx e^{i\omega R/c}$.
The scalar potential is useful for calculating the radiated power, given by
\begin{equation}
    \label{eq-ed-sp}
    V_\text{ed}=-\frac{i\mu_0c\omega\epsilon e}{4\pi R}\left[ e^{i\omega R/c}-I_1^{(\omega)}(R) \right]\hat n\cdot \vec D_e.
\end{equation}
The ``magnetic'' and ``electric'' fields are thus 
\begin{gather}
    \vec H_\text{ed}=\frac{\epsilon e\omega^2}{4\pi cR}\left[ e^{ikR}-I_1^{(\omega)}(R) \right]\hat n\times \vec D_e,\\
    \begin{split}
    \label{eq-ed-ef}
    \vec E_\text{ed}=&\frac{\omega^2\epsilon e}{4\pi\epsilon_0c^2R}\bigg\{ \left[e^{i\omega R/c}-I_2^{(\omega)}(R)\right]\hat n\times(\vec D_e\times\hat n)\\
    &+\left(\frac{\mm c}{\omega}\right)^2 \left[e^{i\omega R/c}-I_0^{(\omega)}\right]\vec D_e\bigg\}.
    \end{split}
\end{gather}
In the above computation, one uses the relation $\pd_jI_n^{(\omega)}(R)\approx i\frac{\omega}{c}\hat n_jI_{n+1}^{(\omega)}(R)$.
These can be used to calculate the Poynting vector $\vec S_\text{ed}$, and then, integrate it over a 2-sphere at a large $R$ to get the radiated  power 
\begin{equation*}
    \label{eq-pow-1}
        \mathcal P_\text{ed}=
                    \frac{\epsilon^2e^2\omega^4}{6\pi\epsilon_0c^3}d_e^2\frac{1+(\mm c/\omega)^2/2}{\sqrt{1-(\mm c/\omega)^2}}\Theta(\omega-\mm c),
\end{equation*}
where the average over the wavelength has been performed.
The total energy of the binary system in GR is \cite{Maggiore:1900zz}
\begin{equation}
    \label{eq-ben}
    E=-\frac{(G\mathcal M\omega)^{2/3}}{2c^2}\mathcal M.
\end{equation}
Now, in this work, $G$ should be replaced by $G'$.
The rate of change in orbital frequency due to the ``electric'' dipole radiation is thus \cite{Hou:2017cjy}
\begin{equation*}
    \label{eq-dfdt-ed}
    \dot f_\text{ed}=\frac{\epsilon^2e^2\omega^3}{4\pi^2\epsilon_0c^3}\eta^{2/5}\mathcal M\xi_e^2\frac{1+(\mm c/\omega)^2/2}{\sqrt{1-(\mm c/\omega)^2}}\Theta(\omega-\mm c).
\end{equation*}
This is the leading order DP correction to the orbital frequency evolution.

Usually in the classical electrodynamics, it is sufficient to consider the dipole radiation \cite{Jackson:1998nia}. 
However, in our discussion, we might need to consider higher order corrections, because for some compact binary system, the dipole radiation vanish. 
For example, consider the DP radiation emitted by a binary neutron star system, then for both stars, the proton mass fractions $\chi_1=\chi_2=0$ and the neutron mass fractions $\chi'_1=\chi'_2=0$. 
So by Eq.~\eqref{eq-eds}, $\xi_e=0$, and  the electric dipole moment $\vec D_e$ vanishes. 
In this case, one has to consider ``magnetic'' dipole and ``electric'' quadrupole radiation. 
As a matter of fact, the calculation in the next subsection shows that ``magnetic''  dipole and ``electric'' quadrupole moments also depend on the sum of proton and neutron mass fractions, so in general, they are nonzero.

\subsection{``Magnetic'' dipole and ``electric'' quadruple radiation}
\label{sec-mdp-eqp}

Similar computation can be done for the ``magnetic'' dipole and the ``electric'' quadrupole radiation.
Following Ref.~\cite{Jackson:1998nia}, the variations of $\vec D_m$ and $Q_{jk}$ induce the following potentials,
\begin{gather*}
    V_\text{md}=0, \quad
    \vec A_\text{md}=\frac{i\mu_0\omega\epsilon e}{4\pi cR}\left[e^{i\omega R/c}-I_2^{(\omega)}(R)\right]\hat n\times \vec D_m,\\
    V_\text{eq}=-\frac{\mu_0\omega^2\epsilon e}{6\pi R}\left[ e^{i2\omega R/c}-I_2^{(2\omega)}(R) \right]\hat n\cdot\vec Q(\hat n),\\
    \vec A_\text{eq}=-\frac{\mu_0\omega^2\epsilon e}{6\pi cR}\left[e^{i2\omega R/c}-I_1^{(2\omega)}(R)\right]\vec{Q}(\hat n),
\end{gather*}
respectively, where the subscripts ``md'' means ``magnetic dipole'' and ``eq'' means ``electric quadrupole''.
Here, the ``magnetic'' dipole and ``electric'' quadrupole moments take the following forms,
\begin{gather*}
    \vec D_m=d_m\hat{\mathscr M},\;d_m=\frac{\mathcal Mc^2}{4\omega}\left( \eta\frac{G'\mathcal M}{c^2}\frac{\omega}{c} \right)^{2/3}\xi_m,\nonumber\\
    \xi_m= \sigma\frac{\Sigma \chi-\sqrt{1-4\eta}\Delta \chi}{m_p}+\frac{\Sigma \chi'-\sqrt{1-4\eta}\Delta \chi'}{m_n},\\
    Q_{jk}=\mathcal Q \hat{\mathscr Q}_{jk},\; \mathcal Q=\frac{3}{\omega}d_m.
\end{gather*}
In these expressions, $\Sigma \chi=\chi_1+\chi_2$, $\Delta\chi=\chi_1-\chi_2$, $\Delta\chi'=\chi'_1-\chi'_2$, $\Sigma \chi'=\chi'_1+\chi'_2$, $\hat{\mathscr M}=(0,0,1)$, and $\hat{\mathscr Q}_{xx}=-\hat{\mathscr Q}_{yy}=-i\hat{\mathscr Q}_{xy}=-i\hat{\mathscr Q}_{yx}=1$ and the remaining components of $\hat{\mathscr Q}_{jk}$ vanish.
Finally, $\vec Q(\hat n)$ has the following components $Q_j(\hat n)=Q_{jk}\hat n^k$.
It is worth to note that $\xi_m$ is also a function of $\Sigma\chi$ and $\Sigma\chi'$, so it is nonvanishing, and ``magnetic'' dipole and ``electric'' quadrupole radiation always exist.
The corresponding ``magnetic'' and ``electric'' fields are
\begin{gather*}
    \vec H_\text{md}=\frac{\omega^2\epsilon e}{4\pi c^2R}\left[ e^{i\omega R/c}-I_3^{(\omega)} \right](\hat n\times\vec D_m)\times\hat n,\\
    \begin{split}
    \vec E_\text{md}=-\frac{\omega^2\epsilon e}{4\pi\epsilon_0c^3R}\bigg[&e^{i\omega R/c}-I_4^{(\omega)}+\frac{\mathfrak m^2c^2}{\omega^2}\times\\
    &\left(e^{i\omega R/c}-I_2^{(\omega)}\right)\bigg]\hat n\times\vec D_m,
    \end{split}\\
    \vec H_\text{eq}=-i\frac{\omega^3\epsilon e}{3\pi c^2R}\left[ e^{i2\omega R/c}-I_2^{(2\omega)} \right]\hat n\times\vec Q(\hat n),\\
    \begin{split}
    \vec E_\text{eq}=-\frac{i\omega^3\epsilon e}{3\pi\epsilon_0c^3R}\bigg[ &\left(e^{i2\omega R/c}-I_3^{(2\omega)}\right)\hat n\times[\vec Q(\hat n)\times \hat n]\\
    &+\left(\frac{\mathfrak mc}{2\omega}\right)^2\left(e^{i2\omega R/c}-I_1^{(2\omega)}\right)\vec{Q}(\hat n)\bigg],
    \end{split}
\end{gather*}
which contribute to radiated power, given by
\begin{gather*}
    \mathcal P_\text{md}=\frac{\omega^4\epsilon^2e^2}{12\pi\epsilon_0c^5}d_m^2\left[ 1- \left( \frac{\mm c}{\omega} \right)^2 \right]^{3/2}\Theta(\omega-\mm c),\\
    \mathcal P_\text{eq}=\frac{\omega^4\epsilon^2e^2}{45\pi\epsilon_0c^5}d_m^2\sqrt{1-\left( \frac{\mm c}{2\omega} \right)^2}\left[ 1+\frac{2}{3}\left( \frac{\mm c}{2\omega} \right)^2 \right]\times\\
    \Theta(2\omega-\mm c).
\end{gather*}
Therefore, the frequency evolves according to 
\begin{gather*}
    \dot f_\text{md}=\frac{\epsilon^2e^2}{128\pi^2\epsilon_0}\left( \frac{\omega}{c} \right)^3\mathcal M\left( \frac{G\mathcal M}{c^2}\frac{\omega}{c} \right)^{2/3}\xi_m^2\times\\
    \left[ 1- \left( \frac{\mm c}{\omega} \right)^2 \right]^{3/2}\Theta(\omega-\mm c),\\
    \dot f_\text{eq}=\frac{\epsilon^2e^2}{480\pi^2\epsilon_0}\left( \frac{\omega}{c} \right)^3\mathcal M\left( \frac{G\mathcal M}{c^2}\frac{\omega}{c} \right)^{2/3}\xi_m^2\times\nonumber\\
    \sqrt{1-\left( \frac{\mm c}{2\omega} \right)^2}\left[ 1+\frac{2}{3}\left( \frac{\mm c}{2\omega} \right)^2 \right]\Theta(2\omega-\mm c),
\end{gather*}
corresponding to $\mathcal P_\text{md}$ and $\mathcal P_\text{eq}$, respectively.

Then, the total change in the orbital frequency evolution due to the DP radiation is 
\begin{equation}
    \label{eq-fevo-dp}
    \dot f_{\gamma'}=\dot f_\text{ed}+\dot f_\text{md}+\dot f_\text{eq},
\end{equation}
omitting higher order contributions.
This would definitely affect the phase evolution of the GW emitted simultaneously \cite{Cardoso:2016olt,Hou:2019jhu}.
However, in the following discussion, one assumes $\dot f_{\gamma'}$ due to the DP radiation is much smaller than that due to the GW given by,  
\begin{equation}
    \label{eq-dfdt-gw}
    \dot f_\text{gw}
    =\frac{48}{5\pi}\left( \frac{c^3}{G'\mathcal M}\right)^2\left(  \frac{G'\mathcal M}{c^2}\frac{\omega}{c}\right)^{11/3}.
\end{equation}
otherwise $\dot f$ would cause large enough GW dephasing which should be observable.
One can check that this assumption implies that the following upper bound is sufficient,
\begin{equation}
    \label{eq-ub1}
\epsilon^2\lesssim10^{-37},
\end{equation}
and later, one ignores the effect of $\dot f_{\gamma'}$.
This requirement also implies that one has only to consider the effect of the DP radiation up to the order of $\epsilon^2$.

The total ``electric'' field strength is $\vec E=\vec E_\text{ed}+\vec E_\text{md}+\vec E_\text{eq}+\cdots$ with the dots representing higher-order corrections.
This field will accelerate mirrors in interferometers as to be discussed below.

\section{Interferometer responses to dark photons}
\label{sec-res}

Let us assume the two mirrors of one arm (labeled by $n=1,2$) of an interferometer are located at $\vec x^{(n)}_a$ and $\vec x^{(n)}_b$ such that $\vec L^{(n)}=\vec x^{(n)}_a-\vec x^{(n)}_b$.
Let $|\vec L^{(n)}|=L$ if there is no dark photon radiation.
These mirrors  move at very small speeds, so the ``Lorentz force'' is basically determined by the ``electric field'' $\vec E$.
Integrating Eq.~\eqref{eq-lf} twice with $\vec B$ ignored, one obtains the change in the position of the mirror $\vec x^{(n)}_a$,
\begin{equation}
    \label{eq-chpos}
\Delta\vec x^{(n)}_a=\frac{\epsilon eN}{m}\int\text dt\int\text dt'\vec E(\vec x^{(n)}_a),
\end{equation}
and a similar expression for the mirror $\vec x^{(n)}_b$. 
So the total change in $\vec L^{(n)}$ is 
\begin{equation}
    \label{eq-tcln}
    \begin{split}
        \Delta \vec L^{(n)}=&\Delta \vec x^{(n)}_a-\Delta\vec x^{(n)}_b\\
        \approx&\frac{\epsilon eN}{m}\int\text dt\int\text dt'\vec L^{(n)}\cdot\nabla\vec E(\vec x^{(n)}_b),
    \end{split}
\end{equation}
where the approximation can be made because the wavelength of the dark photon radiation is on the same order as that of the GW, much larger than $L$.
Since the change in the arm length is $\Delta L^{(n)}=\hat L^{(n)}\cdot \Delta \vec L^{(n)}$ with $\hat L^{(n)}=\vec L^{(n)}/L$, the strain induced by the dark photon radiation is thus 
\begin{equation}
    \label{eq-strain}
    \begin{split}
    h=\frac{\Delta L^{(1)}-\Delta L^{(2)}}{L}
    =\frac{\epsilon eN}{m}D_{jk}\int\text dt\int \text dt'\partial^jE^k,
    \end{split}
\end{equation}
where $D_{jk}=\hat L^{(1)}_j\hat L_k^{(1)}-\hat L_j^{(2)}\hat L_k^{(2)}$ is the detector configuration tensor.
In the following, we set $\chi\approx \chi'\approx 0.5$ for mirrors.
One should realize that in the above derivation, one does not assume the angle between the two arms.
Therefore, Eq.~\eqref{eq-strain} is applicable to ground-based interferometers. 
It might also be useful for space-borne detectors, such as DECIGO/B-DECIGO \cite{Seto:2001qf,Yagi:2011wg,Nakamura:2016hna,Isoyama:2018rjb} and LISA \cite{Seoane:2013qna}.

Now, substituting the ``electric'' field determined in the previous section, one finds the strain $h(t)$ in the time domain, and furthermore, using the stationary-phase approximation \cite{Droz:1999qx,Hou:2019jhu}, one can determine the strain in the frequency domain, given by 
\begin{equation*}
    \begin{split}
    \tilde{h}(f)=&\frac{1}{4}\sqrt{\frac{5\pi}{3}}\frac{\epsilon^2e^2}{4\pi\epsilon_0c^2R}\frac{N}{m}e^{-i\tilde\Phi(\omega)}\bigg\{\frac{\eta^{8/15}}{i}\frac{G\mathcal M^2}{c^3}\\
        &\left( \frac{G\mathcal M}{c^2}\frac{\omega}{c} \right)^{-3/2}\xi_e\left[ \left( 1-e^{-i\omega R/c}I_3^{(\omega)} \right)F_\text{ed}\right.\\
        &\left.+\left( \frac{\mathfrak m_zc}{\omega} \right)^2\left( 1-e^{-i\omega R/c}I_1^{(\omega)}  \right)F'_\text{ed} \right]+i\frac{\eta^{2/3}}{4}\\
        &\frac{G\mathcal M^2}{c^3}\left( \frac{G\mathcal M}{c^2}\frac{\omega}{c} \right)^{-7/6}\xi_m\bigg[1-e^{-i\omega R/c}I_5^{(\omega)}\\
        &+\left( \frac{\mathfrak m_zc}{\omega} \right)^2\left( 1-e^{-i\omega R/c}I_3^{(\omega)} \right)\bigg]F_\text{md}\bigg\}_{\omega=2\pi f}\\
        &+\frac{1}{4}\sqrt{\frac{5\pi}{3}}\frac{\epsilon^2e^2}{4\pi\epsilon_0c^2R}\frac{N}{m}\frac{2\eta^{2/3}G\mathcal M^2}{c^3} \left( \frac{G\mathcal M}{c^2}\right.\\
        &\left.\frac{\omega}{c} \right)^{-7/6}\xi_me^{-i\tilde\Phi'(\omega)}\bigg[\left(1-e^{-i2\omega R/c}I_4^{(2\omega)}\right)\\
        &F_\text{eq}-\left( \frac{\mathfrak m_zc}{2\omega} \right)^2
        \left( 1-e^{-i2\omega R/c}I_2^{(2\omega)} \right)F'_\text{eq}\bigg]_{\omega=\pi f},
    \end{split}
\end{equation*}
where $f$ is the frequency in the Fourier transformation, $F_\text{ed}=D^{jk}\hat n_j[\hat n\times(\hat{\mathscr E}\times\hat n)]_k$, $F'_\text{ed}=D^{jk}\hat n_j\hat{\mathscr E}_k$, $F_\text{md}=D^{jk}\hat n_j[\hat n\times \hat {\mathscr M}]_k$, $F_\text{eq}=D^{jk}\hat n_j\big\{\hat n\times[\hat n\times\hat {\mathscr Q}(\hat n)]\big\}_k$, and $F'_\text{eq}=D^{jk}\hat n_j\hat {\mathscr Q}_k(\hat n)$.
$R$ has to be replaced by the luminosity distance $d_L$, and all the quantities on the right-hand side should be understood as the measured ones in the detector frame.
In particular, $\mm_z=\mm(1+z)$ is the redshifted DP mass with $z$ the redshift of the source.
In addition, $\tilde\Phi(\omega)=\omega t_c-\Phi_c-\frac{\pi}{4}+\frac{3}{256}\left( \frac{G\mathcal M}{c^2}\frac{\omega}{c} \right)^{-5/3}$, and $\tilde\Phi'(\omega)=2\tilde\Phi(\omega)+\frac{\pi}{4}$ with $t_c$ and $\Phi_c$ the fiducial coalescence time and (orbital) phase, respectively.

The signal-to-noise ratio $\rho$ can be calculated from \cite{Yunes:2013dva}
\begin{equation}
    \label{eq-snr}
    \rho^2= 4\int\frac{\langle|\tilde h(f)|^2\rangle}{S_n(f)}\ud f,
\end{equation}
where $\langle\rangle$ means to take the angular average, as the angular resolution of a single laser interferometer is still poor.
By requiring $\rho<8$, one can determine the bounds on $\epsilon^2$ and $m_{\gamma'}$, since no dark photon radiation has been observed yet.

Before presenting the constraints, one should note the presence of $\mm_z$ in the above waveform. 
If DPs exist and $\mm$ can be measured independently and locally, the above waveform explicitly depends on the source redshift $z$.
Thus, $z$ can be determined directly from the DP induced waveform, and this has a great significance in cosmology, especially resolving the Hubble tension \cite{Riess:2019cxk,Aghanim:2018eyx}.

\section{Constraints}
\label{sec-cts}

We will determine the constraints from the absence of DP signals in some of the observed GW events by aLIGO, and by assuming other interferometers (e.g., ET, CE, DECIGO, B-DECIGO and LISA, et.~al.) cannot detect DPs either.
To compute SNR, one needs to set the integration limits in Eq.~\eqref{eq-snr}.
We will set the lower limit $f_\text{lower}$ to be 1 Hz for ET, 5 Hz for aLIGO and CE, $10^{-3}$ Hz for DECIGO and B-DECIGO, and $10^{-4}$ Hz for LISA.
The upper limit will be $f_\text{upper}=\text{min}\{f_\text{isco},f_\text{max}\}$, where $f_\text{isco}$ is the frequency corresponding to the inner-most stable circular oribt \cite{Bonvin:2016qxr}, 
\begin{equation*}
  f_\text{isco}=4.40(1+1.25\eta+1.08\eta^2)\left[ \frac{M_\odot}{(1+z)(m_1+m_2)} \right]\text{kHz},
\end{equation*}
and $f_\text{max}$ is the upper bound of the detector sensitivity band.
For ground-based interferometers, $f_\text{max}=10^4$ Hz; 
for DECIGO and B-DECIGO, $f_\text{max}=100$ Hz, and $f_\text{max}=1$ Hz, if LISA is used.

\begin{figure}[t]
    \centering
    \includegraphics[width=0.45\textwidth]{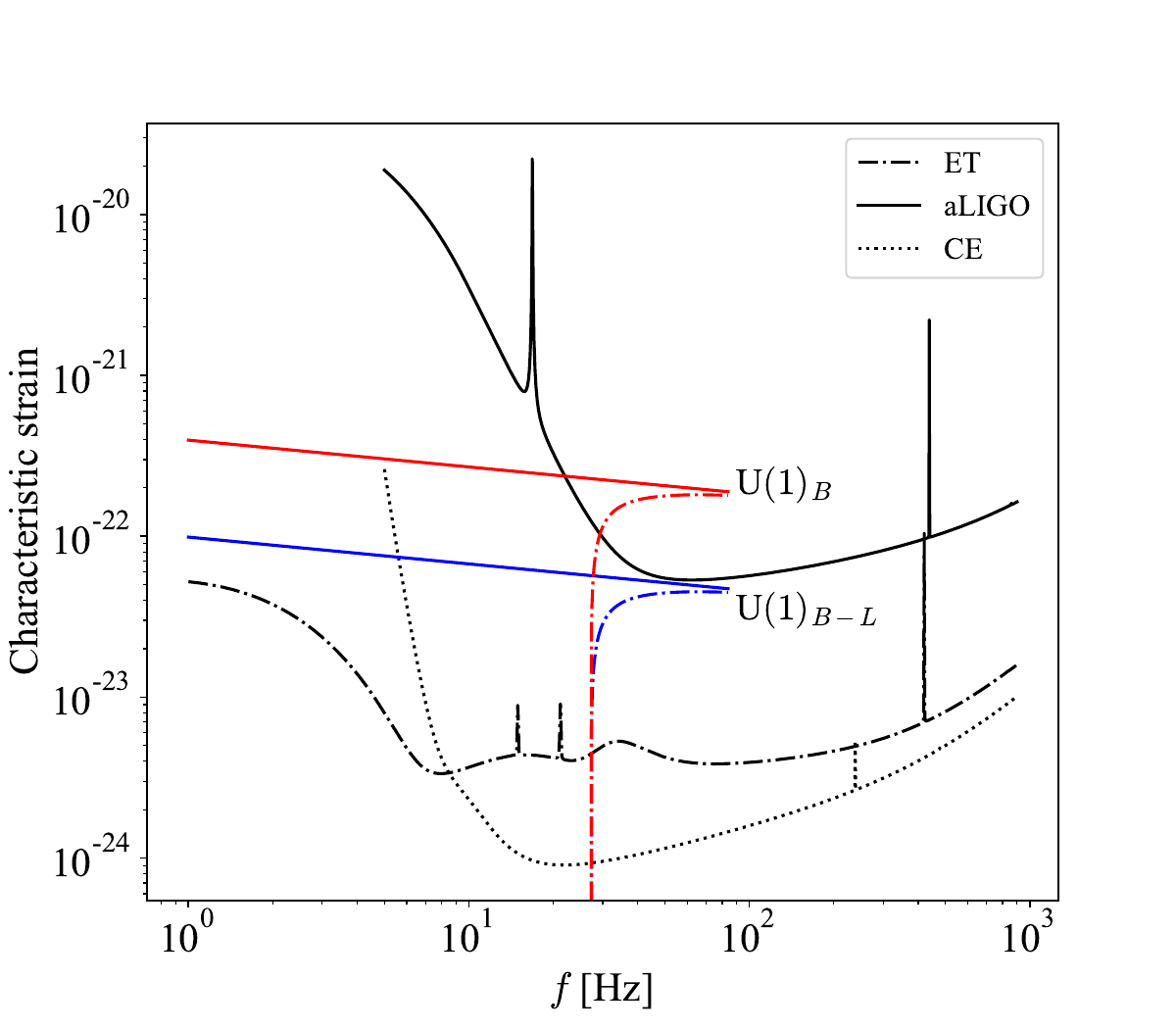}
    \includegraphics[width=0.45\textwidth]{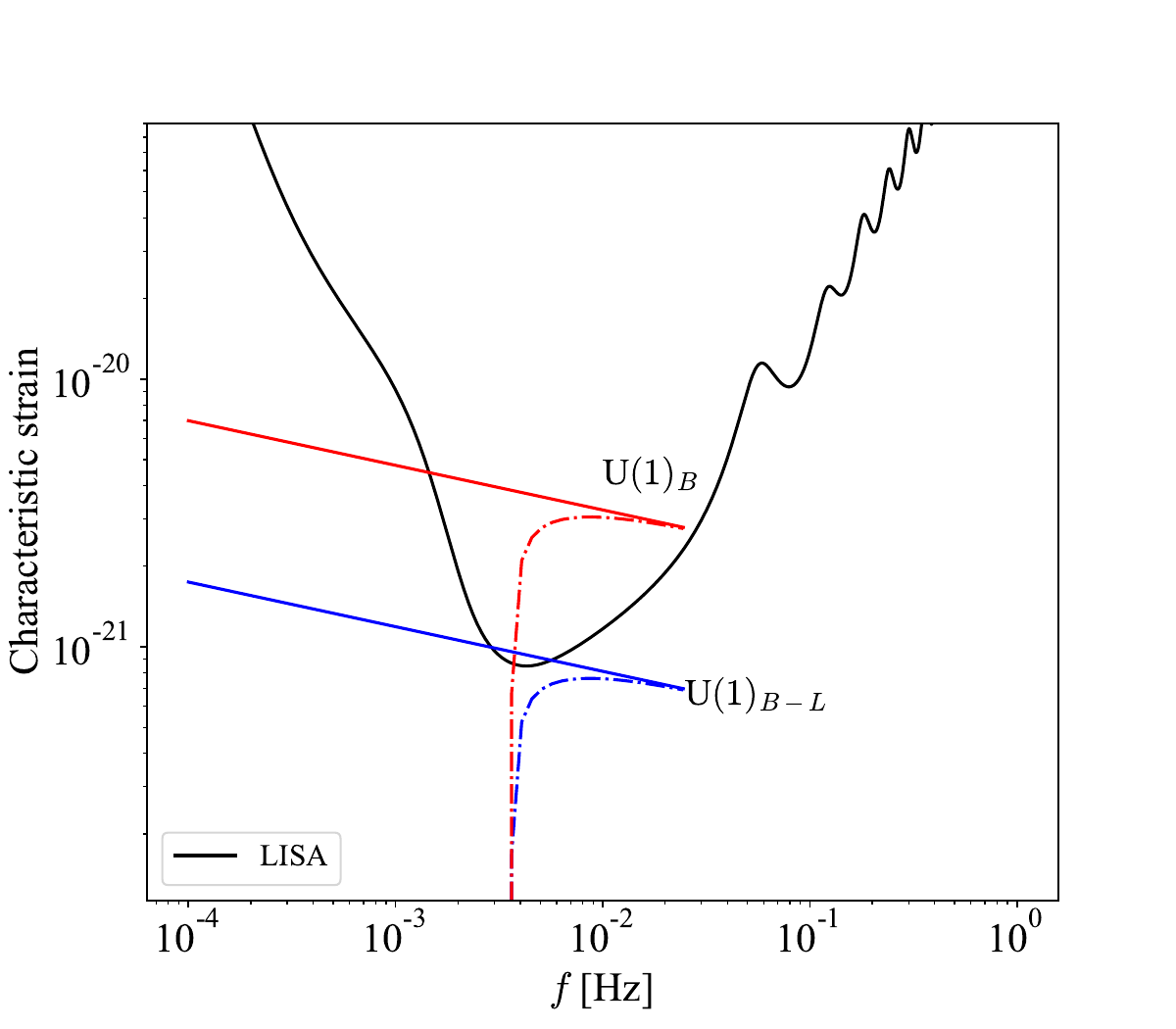}
    \caption{(a) The upper panel: The characteristic strains induced by the DP radiation generated by a GW150914-like binary system.
    The black solid, dot-dashed and dotted curves are the sensitivities for aLIGO, ET and CE, respectively.
    The green and cyan solid curves are the sensitivities for DECIGO and B-DECIGO, respectively.
    The red and blue curve are signals of DP radiation for $U(1)_B$ and $U(1)_{B-L}$, respectively, with the solid ones for $m_{\gamma'}=0$ and the dot-dashed ones for $m_{\gamma'}\ne0$.
    Here, one chooses $m_{\gamma'}c^2\approx10^{-13}$ eV for the purpose of demonstration. 
    This mass corresponds to the Compton frequency $25$ Hz.
    (b) The lower panel: The characteristic strains induced by the DP radiation generated by an IMRI.
    The black curve is the sensitivity for LISA.
    The mass of DP in this case is chosen to be $1.5\times10^{-17}$ eV, corresponding to $2\times10^{-3}$ Hz.
    Drew with PyCBC \cite{pycbc}.}
    \label{fig-dps}
\end{figure}

Now, let us first demonstrate some examples of waveform of the DP radiation.
Figure \ref{fig-dps} displays the characteristic strains induced by the DP radiation generated by GW150914 and IMRI, assuming $\epsilon^2$ takes the upper bound set by Eq.~\eqref{eq-ub1}.
In the upper panel, the black curves are the sensitivity curves for aLIGO (solid), ET (dot-dashed) and CE (dotted), and the green and cyan solid curves are for DECIGO and B-DECIGO, and the remaining curves are the DP signals with the solid curves for the massless case, and the dot-dashed ones for the massive case.
Since for the massive case, the radiation would be shut down if the frequency is too low, the dot-dashed curves start from $f=25$ Hz, which corresponds to $m_{\gamma'}c^2\approx 10^{-13}$ eV.
In the lower panel, we display the DP signals (labeled in the same way as in the upper panel) for LISA (the black solid curve).
Here, the DP mass $m_{\gamma'}c^2\approx1.5\times10^{-17}$ eV and the Compton frequency is $2\times10^{-3}$ Hz for the purpose of demonstration.
In drawing these figures, one has already taken the angular averages.

So now, the constraints on DP model are presented assuming $\rho<8$ for DP radiation.
If the DP is massless, the upper bounds on $\epsilon^2$ are listed in Table~\ref{tab-epbml}.
Numbers enclosed by brackets are for $U(1)_{B-L}$, and these not enclosed are for $U(1)_B$.
\begin{table}[h]
    \centering
    \begin{tabular}{c|cccc}
        \hline\hline
        Detector & GW150914 & GW170817 & GW200105 & \\
        \hline
        aLIGO & 1.9(7.6) & 2.2(4.4) & 6.8(13.3) & \\
        ET-D & 0.08(0.32) & 0.11(0.23) & 0.34(0.58) & \\
        CE & 0.02(0.09) & 0.03(0.07) & 0.10(0.18) &\\
        DECGIO & 0.02(0.07) & 0.03(0.05) & 0.08(0.05) & \\
        B-DECIGO & 0.21(0.86) & 0.34(0.68) & 1.0(0.78) & \\
        \hline
        & EMRI & IMRI & IMBH & SMBH  \\
        \hline
        LISA & 19.2(76.8) & 1.5(6.1) & 1.55(6.2) & 7.5(30.0)  \\
        \hline\hline
    \end{tabular}
    \caption{The upper bounds (in units of $10^{-37}$) on $\epsilon^2$ assuming $m_{\gamma'}=0$.
    }
    \label{tab-epbml}
\end{table}
From this table, one knows that aLIGO and LISA are incapable of putting stronger constraints than Eq.~\eqref{eq-ub1}.

\begin{figure}[t]
    \centering
    \includegraphics[width=0.45\textwidth]{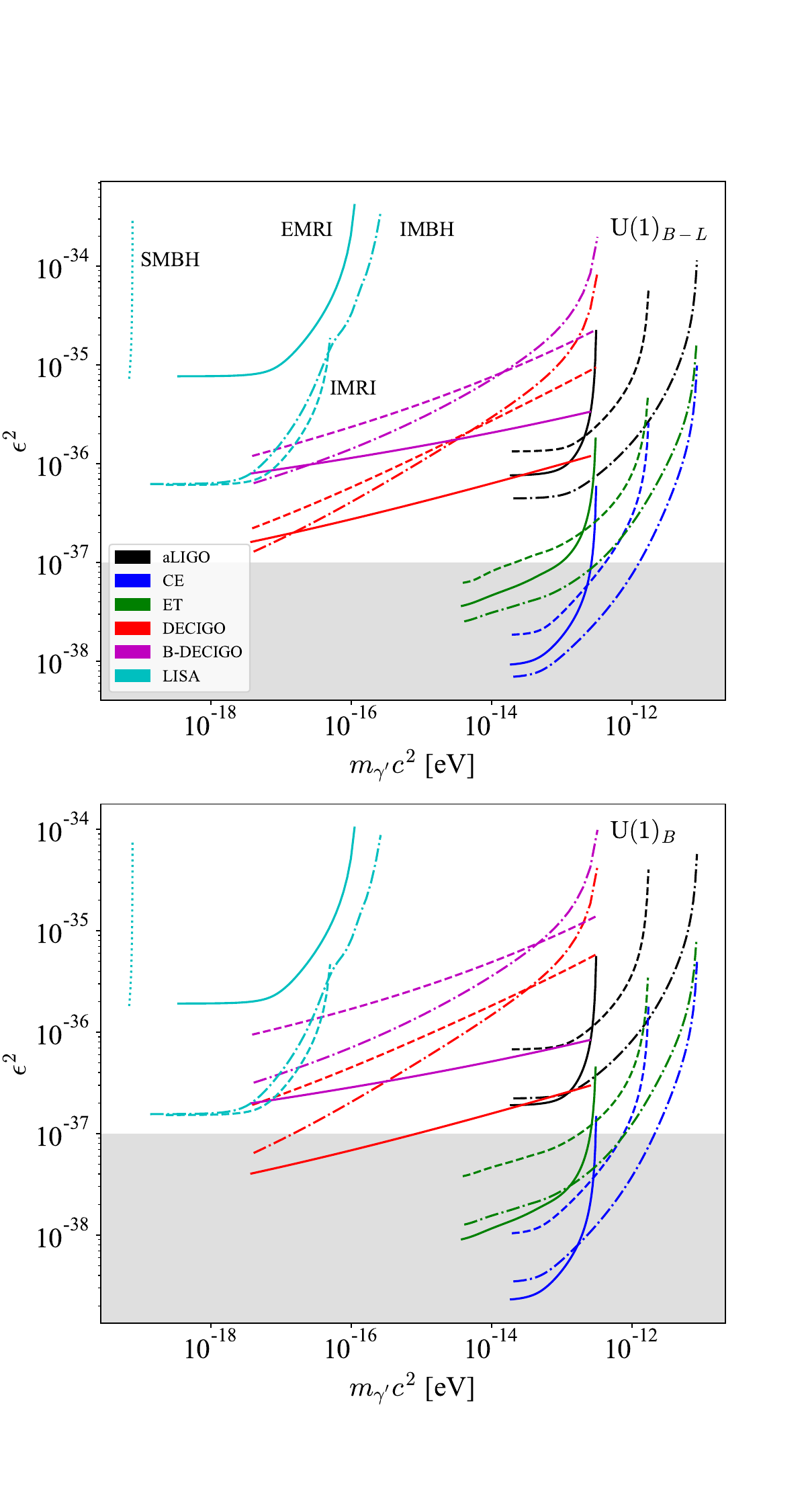}
    \caption{Constraints on $m_{\gamma'}$ and $\epsilon^2$ derived from the absence of the DP signals in several interferometers.
    The upper panel shows the constraints for $U(1)_{B-L}$ and upper for $U(1)_B$.
    The shaded areas are due to Eq.~\eqref{eq-ub1}.
    The constraints derived from the future observations by LISA are represented by the cyan curves, as clearly labeled.
    The remaining curves are for other detectors.
Among them, the solid curves are from GW150914, the dashed ones from GW200105, and the dot-dashed ones from GW170817.
Drew with PyCBC \cite{pycbc} and LISA sensitivity calculator \cite{neil_cornish_2019_2636514}.}
    \label{fig-cs}
\end{figure}

If $m_{\gamma'}\ne0$, the constraints are shown in Fig.~\ref{fig-cs}.
The upper panel displays the constraints for $U(1)_{B-L}$, and the lower panel is for $U(1)_{B}$.
The shaded areas corresponding to the bound Eq.~\eqref{eq-ub1}.
The constraints derived from the future observations by LISA are represented by the cyan curves, as clearly labeled.
The remaining curves are constraints for other detectors, as indicated by different colors. 
Among them, the solid curves are from GW150914, the dashed ones from GW200105, and the dot-dashed ones from GW170817.
All the constraints on $\epsilon^2$ in the lower panel are less than those in the upper panel.
This figure also shows that CE and ET impose stronger constraints on $\epsilon^2$, while the remaining detectors mainly provide even less stringent bounds than Eq.~\eqref{eq-ub1}.

\section{Conclusion}
\label{sec-con}

In this work, the (massive) DP radiation emitted by orbiting binary stars is computed for the first time.
Its waveform up to the ``electric'' quadrupole radiation is obtained, which depends on the source redshift explicitly.
Then, the response of the laser interferometer to the DP radiation is determined, valid for all possible configurations of the two arms.
Since no obvious deviations from GR's prediction have been detected in the GW strain by LIGO/Virgo, three types of constraints can be applied to the DP model: 1) the effective gravitational constant $G'\approx G$; 2) the orbit of the binary system decays approximately due to the GW emission; 3) the SNR for the DP signal should be small.
These requirements lead to constraints collected in Table~\ref{tab-epbml} for massless DPs and Fig.~\ref{fig-cs} for massive DPs.
Although these constraints are weaker than those reported in Refs.~\cite{Pierce2018.PRL.121.061102,Guo:2019ker,Morisaki2021.PRD.103.L051702,LIGOScientific:2021odm}, they are the first constraints derived from the DP radiation without the assumption of DP being DM.
Of course, in the current work, we have considered the DP radiation only in the inspiral stage of the coalescence of the binary system.
Since the SNR for the GW mainly comes from the GW signal produced during the merger and ring-down stages, then if one also studies the DP radiation during these stages, one should obtain stronger constraints.

The method presented in this work is actually applicable  to many other elementary particles in new physics, including other DM candidates, such as axion \cite{Peccei:1977ur,Kim:1979if,Shifman:1979if,Dine:1981rt}.
As long as these particles interact with the visible matter, they can be produced in processes such as the coalescence of binary systems, the spinning of neutron stars with mountains \cite{Ushomirsky:2000ax,Horowitz:2009ya}, and the phase transitions in the very early universe \cite{Hasegawa:2019amx}.
Once they reach the interferometer, they induce new strains in addition to that due to the GW.
It is also possible to use pulsar timing arrays to detect DPs and other particles in new physics \cite{Dror:2021wrl,Xue:2021xts}.
In fact, pulsar timing arrays detect the frequency shift $\Delta f=f_r-f_e$ of photons, where $f_r=-k_\mu u^\mu_\otimes$ is the photon frequency measured by an observer with 4-velocity $u^\mu_\otimes$ on the earth, and $f_e=-k_\mu u^\mu_p$ measured by an observer with 4-velocity $u^\mu_p$ on the pulsar  \cite{1975GReGr...6..439E,1978SvA....22...36S,Detweiler:1979wn,Hou:2017bqj}.
$k^\mu$ is the photon 4-velocity.
When there is no stochastic GW background or DP radiation background, $\Delta f=0$.
If the stochastic GW background exists, the physical distance between a pulsar and the earth is changing, so the relative velocity between them is nonzero, which results in $\Delta f\ne0$.
Due to the stochastic nature of the GW background, $\Delta f$'s of photons coming from different directions are correlated, as described by the famous Hellings-Downs curve \cite{Hellings:1983fr}.
Similarly, if the stochastic DP radiation background exists, both the pulsar and the earth interact with DPs.
Then there exists relative velocity, and $\Delta f\ne0$, too.
The correlation of $\Delta f$ caused by DPs is expected to be different from the Hellings-Downs curve, as suggested by the correlations due to the vector polarizations in some modified theories of gravity \cite{Gong:2018cgj,*Hou:2018djz,*Gong:2018vbo}.
Of course, as in Ref.~\cite{Dror:2021wrl}, the emission of DPs also changes the orbital decay rate of the binary system, modifying the spectrum of the stochastic GW background, which might be detected by pulsar timing arrays.
As a matter of fact, it might be better to use pulsar timing arrays to detect DPs, as their sensitivity band is from $10^{-10}$ Hz to $10^{-6}$ Hz.
So DPs of even smaller masses than these considered here will be emitted by binary systems and detected by pulsar timing arrays.
Since DPs are less massive, it would take longer time for black holes to discharge, and so, it is expected that pulsar timing arrays would constrain DP models more strongly.
Although pulsar timing arrays can detect DPs, we will not discuss this possibility in the current work and consider it in future.
Therefore, GW laser interferometers and pulsar timing arrays also serve as tools to detect new physics.

\begin{acknowledgements}
This work was supported by the National Natural Science Foundation of China under Grants No.~11633001, No.~11673008, No.~11922303, and No.~11920101003 and the Strategic Priority Research Program of the Chinese Academy of Sciences, Grant No. XDB23000000.
SH was supported by Project funded by China Postdoctoral Science Foundation (No.~2020M672400).
ST was supported by the Initiative Postdocs Supporting Program under Grant No.~BX20200065 and China Postdoctoral Science Foundation under Grant No. 2021M700481.
\end{acknowledgements}


%

\end{document}